# OVERVIEW OF THE APT ACCELERATOR DESIGN*

J. F. Tooker, R. Bourque, D. Christiansen, J. Kamperschroer, G. Laughon, M. McCarthy,
M. Schulze, General Atomics, San Diego, CA 92186

*Abstract*

The accelerator for the APT Project is a 100 mA CW proton linac with an output energy of 1030 MeV. A High Energy Beam Transport (HEBT) conveys the beam to a raster expander, that provides a large rectangular distribution at a target/blanket (T/B) assembly. Spallation neutrons generated by the proton beam in the T/B reacts with Helium-3 to produce tritium. The design of the APT linac is an integrated normal-conducting (NC)/superconducting (SC) proton linac; the machine architecture has been discussed elsewhere [1]. The NC linac consists of a 75 keV injector, a 6.7-MeV 350-MHz RFQ (radio frequency quadrupole), a 96-MeV 700-MHz CCDTL (coupled-cavity drift-tube linac), and a 700-MHz CCL (coupled-cavity linac), with an output energy of 211 MeV. This is followed by a SC linac, that employs 700-MHz elliptical niobium 5-cell cavities to accelerate the beam to the final energy. The SC linac has two sections, optimized for beam velocities of $\beta = 0.64$ and $\beta = 0.82$. Each section is made up of cryomodules containing two, three, or four 5-cell cavities, driven by 1-MW 700-MHz klystrons. The singlet FODO lattice in the NC linac transitions to a doublet focusing lattice in the SC linac, with conventional quadrupole magnets in the warm inter-module spaces. This doublet lattice is continued in the HEBT. An overview of the current linac design will be presented.

## 1 INTRODUCTION

Figure 1 shows the architecture of the APT 1030 MeV accelerator. It is a normal conducting accelerator up to 211 MeV. A 75-KeV injector generates the cw proton beam and acceleration continues through a RFQ, a CCDTL, and a CCL to 212 MeV. This is followed by a superconducting accelerator to the final energy. The first SC section has 102 five-cell medium-$\beta$ niobium cavities optimized for $\beta = 0.64$. At the end of this section, the proton beam energy is 471 MeV. The second section is a high-$\beta$ section that has five-cell niobium cavities optimized for $\beta = 0.82$. To attain an output energy of 1030 MeV, 140 cavities are needed. A HEBT directs the beam around a 90 degree bend to a 45 degree switchyard, where the beam goes straight to a beam stop for tuning or to the target/blanket.

*Supported by DOE Contract DE-AC04-96AL99607

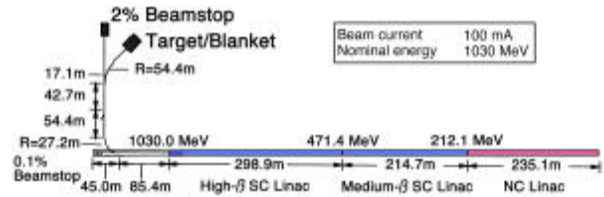

Figure 1: Architecture for APT 1030 MeV accelerator

## 2 ACCELERATOR DESIGN

The following sections describe the design of the various stages of the APT Linac.

### 2.1 Low Energy Linac

The LE linac consists of the injector, a 350-MHz RFQ, a 700-MHz CCDTL, and a 700-MHz CCL.

### 2.1.1 Injector

The 2.8-m long injector has a radio frequency (RF)-driven ion source that produces a 110-mA cw proton beam at 75 keV [1]. A low energy beam transport that has two solenoid magnets matches the proton beam into the acceptance of the RFQ.

### 2.1.2 RFQ

The RFQ is an eight-meter long structure built of four resonantly coupled segments tuned for 350 MHz [1]. The RFQ accepts the 75-keV, 110-mA beam from the injector and produces a 6.7-MeV, 100-mA beam. It is driven by three 1.2-MW klystrons. The RFQ has been operated in the Low Energy Demonstration Accelerator of APT and its performance is discussed elsewhere [2,3].

### 2.1.3 CCDTL

A 700-MHz CCDTL [4] accepts the 100-mA beam from the RFQ and accelerates it to 96 MeV. The CCDTL cavities are grouped into six resonant structures called supermodules that span a length of 112.8 m. The first supermodule consists of a series of side-coupled 2-gap drift-tube linac (DTL) cavities with the quadrupole magnets [5] of the FODO lattice located between them. The focusing lattice begins with a period of 8-$\beta\lambda$ to match the beam from the RFQ and transitions to 9-$\beta\lambda$ at 9 MeV to provide additional space for the quadrupole magnets and beam diagnostics. Module two is made up of DTL cavities with two drift tubes, forming a series of three-gap cavities connected by coupling cells. Modules 3 to 6 consist of two-cavity, two-gap segments. To maintain strong transverse focusing, the quadrupole magnets in the FODO lattice continue with the same 9-$\beta\lambda$ periodicity. The first module is energised by a single one MW klystron. The other five supermodules are energised by up to five klystrons.

### 2.1.4 CCL

The CCL is composed of five supermodules spanning a length of 110.4 m. Each is made up of a string of seven-cell segments side-coupled to form a single resonant structure energised by up to seven one MW klystrons. The singlet 9-$\beta\lambda$ FODO lattice is continued throughout the CCL

The Coupled-Cavity Tuning (CCT) code [6] has been developed to help design the CCL cavities (and later the CCDTL cavities). This code automates the RF calculations using CCLFISH and iterates the geometry of the accelerating cavity, the geometry of the coupling cavity, and the separation between them to achieve the correct frequencies and coupling. The resultant geometry can then be fed into a CAD program to generate the drawings. Four cold models are planned along the length of the CCL to validate the code. Figure 2 shows one of the CCL cold models designed with CCT. Three-D RF analysis and coupled RF/structural analyses are also being performed on this cw RF structure [7] to address cavity sizing and the effects of high RF power densities.

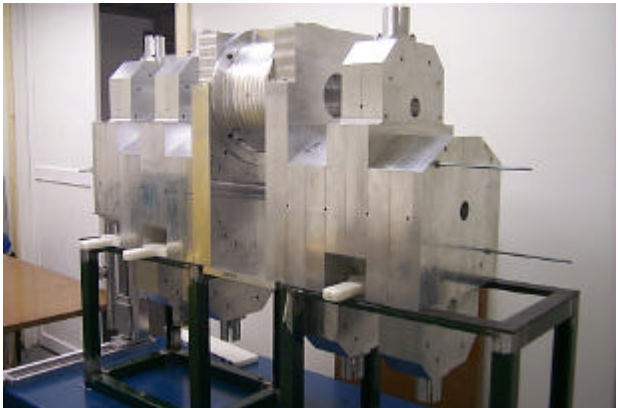

Figure 2: Cold Model of CCL Segments 283 & 284

## 2.2 High Energy Linac

The superconducting cavities of the HE linac are contained in cryomodules that provide the thermal insulation and connection to the cryogenics system to maintain the cavities at their operating temperature of 2.15 K. In the HE linac, the FODO lattice of the LE linac transitions to a doublet lattice, consisting of normal-conducting quadrupole magnets [8] located in the warm regions between the cryomodules. The first six cryomodules in the medium-$\beta$ section contain two cavities each (see Figure 3), providing a shorter 4.877-m focusing period. This was done to improve the match

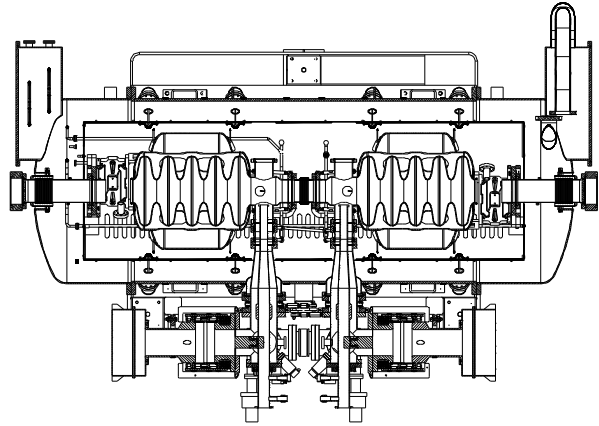

from the LE linac,

Figure 3: Cross-section of two-cavity cryomodule.

The remaining 30 cryomodules in the medium-$\beta$ section contain three cavities each, with a longer 6.181-m focusing period. Each cryomodule in the medium-$\beta$ section is powered by a single 1-MW klystron. Each cavity is fed by a single RF coaxial power coupler, so that the RF power from each klystron is divided by two or three, with up to 420 kW per coupler.

The high-$\beta$ section is a series of 35 four-cavity cryomodules with the room temperature doublet quadrupole magnets in the warm regions between them. This section has a period of 8.540 m. Each high-$\beta$ cryomodule is powered by two 700 MHz klystrons. The power from each klystron is split by two to feed a single RF coaxial power coupler on each of two cavities.

## 2.3 High Energy Beam Transport

The doublet lattice of the HE linac is continued along the transport line of the HEBT. There are ten periods (85.4 m) prior to a 90 degree bend. If the dipole magnets of this bend are de-energized, the beam goes straight to a 0.1% duty cycle beam stop for tuning. If they are energized, the beam is then directed to a 45 degree switchyard. Here the beam can go straight to a beam stop capable of handling 2% of the full beam power. It is used during commissioning, start up, and tuning of the accelerator. The 45° bend in the switchyard can then direct the beam from this beam stop to the target/blanket assembly. The beam is expanded onto the target by a beam raster system [9], which sweeps the beam uniformly over the 19-cm wide by 190-cm high tungsten target.

## 2.4 RF Power System

Three 1.2-MW, 350-MHz klystrons are used to energise the RFQ. Only two are required to accelerate the beam in the RFQ; the third is a spare so that the linac can continue to operate if one of the 350-MHz RF power systems fails. The power from each klystron is split by two, feeding six iris couplers in the RFQ cavity. This is shown in Figure 4.

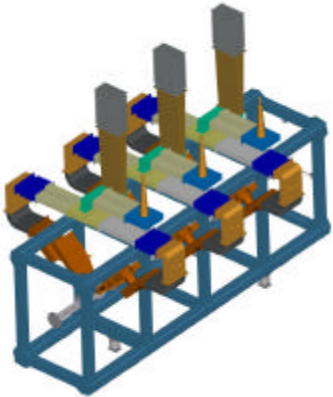

Figure 4: Three waveguide feeds to RFQ Cavity.

The supermodules in the LE linac and the superconducting cavities in the HE linac are powered by 1-MW, 700-MHz klystrons. There are 52 klystrons in the LE linac, 36 klystrons in the medium-$\beta$ section of the HE linac, and 70 klystrons in the high-$\beta$ section of the HE linac for the 1030 MeV accelerator. The power from each klystron is split by two, except for two locations. The first module of the CCDTL is fed by a single klystron, where the power is split by four. For the three-cavity cryomodule, the power is split by three, with each cavity fed by a single power coupler, as shown in Figure 5.

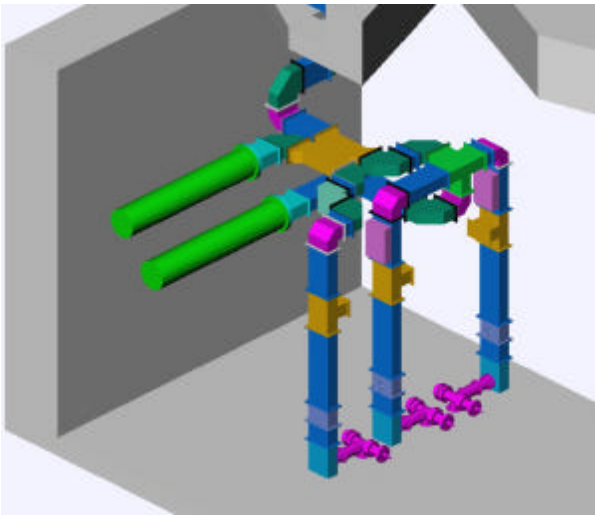

Figure 5: RF power splitting to 3-cavity cryomodule.

## 2.5 Cryogenics System

The cryogenic system [10] supplies helium cooling to maintain the niobium cavities at 2.15 K. This system provides to the 1030 MeV accelerator approximately 15 kW of refrigeration at 2.15 K for the superconducting RF cavities and approximately 64 kW of refrigeration between 5 K and 30 K for the thermal intercepts on the RF power couplers, thermal shields in the cryomodules, and the cryo-distribution system. The cryogenic distribution line contains two sets of supply and return lines, and runs the length of the HE linac. Sets of four U-tube transfer lines connect each cryomodule to this distribution line. The distribution line is supplied by a cryogenics plant, where three, semi-independent helium refrigerators provide the closed-loop helium cooling. Each refrigerator contains a 4-K coldbox using gas-bearing turbine expanders, a 2-K coldbox with cold compressors to generate the sub-atmospheric conditions within the cryomodule, warm helium compressors for gas compression, liquid and gas storage, and appropriate instrumentation and controls.

## REFERENCES

[1] J. Tooker, et al., "Overview of the APT Accelerator Design," Proc. 1999 Particle Accelerator Conf., New York City, March-April 1999
[2] L.M. Young, et. al., "High Power Operations of LEDA," these proceedings.
[3] M. Schulze, et. al., "Beam Emittance Measurements of the LEDA RFQ," these proceedings.
[4] R. Wood, et. al, "Status of Engineering Development of CCDTL for Accelerator Production of Tritium," Proc. 1998 Int. Linac Conf., Chicago (August 1998).
[5] S. Sheynin, et. al, "APT High Energy Linac Intertank Assembly Design," Proc. 1999 Particle Accelerator Conf., New York City, 1999
[6] P.D. Smith, "CCT, a Code to Automate the Design of Coupled Cavities," these proceedings.
{7} G. Spalek, et. al., "Studies of CCL Structures with 3D Codes," these proceedings.
[8] S. Sheynin, "APT High Energy Linac Intertank Assembly Design," Proc. 1999 Particle Accelerator Conf., New York City, 1999.
[9] S. Chapelle, et. al, "Testing of a Raster Magnet System for Expanding the APT Proton Beam," Proc. 1999 Particle Accelerator Conf., New York City, 1999
[10] G. Laughon, "APT Cryogenic System," Proc. of the 1999 Cryogenic Engineering Conf., Montreal, Canada, July, 1999

.